**Topological phonons in an inhomogeneously strained silicon-1: Evidence of long-distance spin transport and unidirectional magnetoresistance of phonons**

Anand Katailiha[1‡], Ravindra G. Bhardwaj[1‡], Paul C. Lou[1‡], Ward P. Beyermann[2], and Sandeep Kumar[1,*]

[1]Department of Mechanical engineering, University of California, Riverside, CA 92521, USA

[2] Department of Physics and Astronomy, University of California, Riverside, CA 92521, USA

[*] Corresponding author

Email: sandeep.suk191@gmail.com

Abstract

Transverse acoustic waves in an inhomogeneous medium are analogues to electromagnetic waves and will exhibit topological behavior due to the Berry gauge potential in the momentum space due to inhomogeneity. The inhomogeneous (or gradient) medium can be created using an applied strain gradient in a semiconductor thin film (silicon) since the phonon frequency and dispersion will be a function of the local strain along the strain gradient direction. As a consequence, topological phonon mediated spin and heat transport can be engineered in the semiconductor thin films. Here, we present evidence of a long-distance (100 µm) spin transport in the freestanding Si thin film sample under an applied strain gradient using transverse spin-Nernst effect measurement. The long-distance spin transport was attributed to the topological spin-Hall effect of phonons in an inhomogeneous medium. The inhomogeneous medium was validated using unidirectional magnetoresistance of phonons where the magnitude of the coefficient of the non-reciprocal response at room temperature was as large as reported in the BiTeBr at low temperatures. The topological phonons also manifested the topological Nernst effect. This work not only enhances the current understanding of inhomogeneous systems but also lays the foundation of the topological and spin phononics.

Keyword- topological phonon, non-reciprocal transport, inhomogeneous strain, silicon, topological Nernst effect, spin transport.

The long-distance spin transport is an essential requirement for realization of spintronics devices. The spin transport in metals and semiconductors relies on electron diffusion, which makes long-distance spin transport difficult, especially at room temperature[1]. This led to research in ferromagnetic and antiferromagnetic insulators where spin transport mediated by magnon diffusion[2,3] at tens of micrometer scale has been reported[4-7]. Recently, Rückriegel and Duine[8] theoretically hypothesized that magnetoelastic coupling at the interface between ferromagnet and non-magnetic insulator materials can lead to spin transport at millimeter scale. The spin transport is expected to be mediated by the angular momentum of the circularly polarized transverse phonons. While long distance spin transport was hypothesized for insulators but any combination of material having magnetoelastic coupling and angular momentum of phonons could be suitable. However, such a long-range spin transport has not been experimentally realized so far.

The angular momentum of the phonons is an essential requirement for long-distance spin transport, which can arise due to the dynamical multiferroicity[9-11]. In a recent study, even centrosymmetric highly doped (conducting) Si thin films were found to exhibit dynamical multiferroicity and temporal magnetic moment under an applied strain gradient[12]. The dynamical multiferroicity in Si can be described as:

$$\boldsymbol{M}_t \propto \boldsymbol{P}_{FE} \times \partial_t \boldsymbol{P} \qquad (1)$$

where $\boldsymbol{M}_t$, $\boldsymbol{P}_{FE}$ and $\boldsymbol{P}$ are temporal magnetic moment, flexoelectronic effect and time dependent polarization of optical phonons, respectively. In addition, the strain gradient also creates a gradient (inhomogeneous) medium in Si thin film since the phonon dispersion and the frequency are a function of position and local strain as shown in Figure 1 (a); similar to gradient index medium for optics[13]. Bliokh and Freilikher[14] theoretically demonstrated that transverse acoustic waves in an inhomogeneous medium were analogous to electromagnetic waves. As a consequence, the deflection of phonon (ray) in an inhomogeneous medium is given by:

$$\delta \boldsymbol{r}_{tc} = -\sigma_c \lambda_{t0} \int_C \frac{\boldsymbol{p}_t \times d\boldsymbol{p}_t}{p_t^3} = -\sigma_c \lambda_{t0} \frac{\partial \Theta^B}{\partial \boldsymbol{p}_{tc}^{(0)}} \qquad (2)$$

where $\sigma_c$, $\lambda_{t0}$, $\boldsymbol{p}_t$ and $\Theta^B$ are helicity, wavelength, momentum and Berry phase[14]. Hence, transverse acoustic waves will give rise to topological spin-Hall effect of phonons similar to the spin-Hall effect of light[13] in the gradient index medium. Consequently, the topological phonons having large magnetic moment due to topological electronic magnetism of phonons can lead to long-distance spin transport in Si.

The experimental verification of the topological spin-Hall effect of phonons in Si can be done with the spin-Nernst effect (SNE) [15,16], which is a thermal counterpart of electronic spin-Hall effect. In the SNE measurement, an angle dependent magneto-thermopower response exhibit modulation from spin current backflow ($J_s^{back}$) due to relative orientation of ferromagnetic moment and spin accumulation as shown in Figure 1 (b). In our experimental setup, a transverse SNE (TSNE) response was measured at away (longer than spin diffusion length) from the heat source, which allowed us to extract spin-transport distance information. Here, we present possible experimental evidence of topological phonon transport in a freestanding Si thin film structure. We experimentally measured a large TSNE response at a distance of 100 µm from the thermal source. The topological spin-Hall effect of phonons was expected to be the underlying cause of long-distance spin transport. The topological phonons arose due to the strain gradient mediated inhomogeneity. The topological Nernst effect and non-reciprocal charge transport response was used to support our hypothesis.

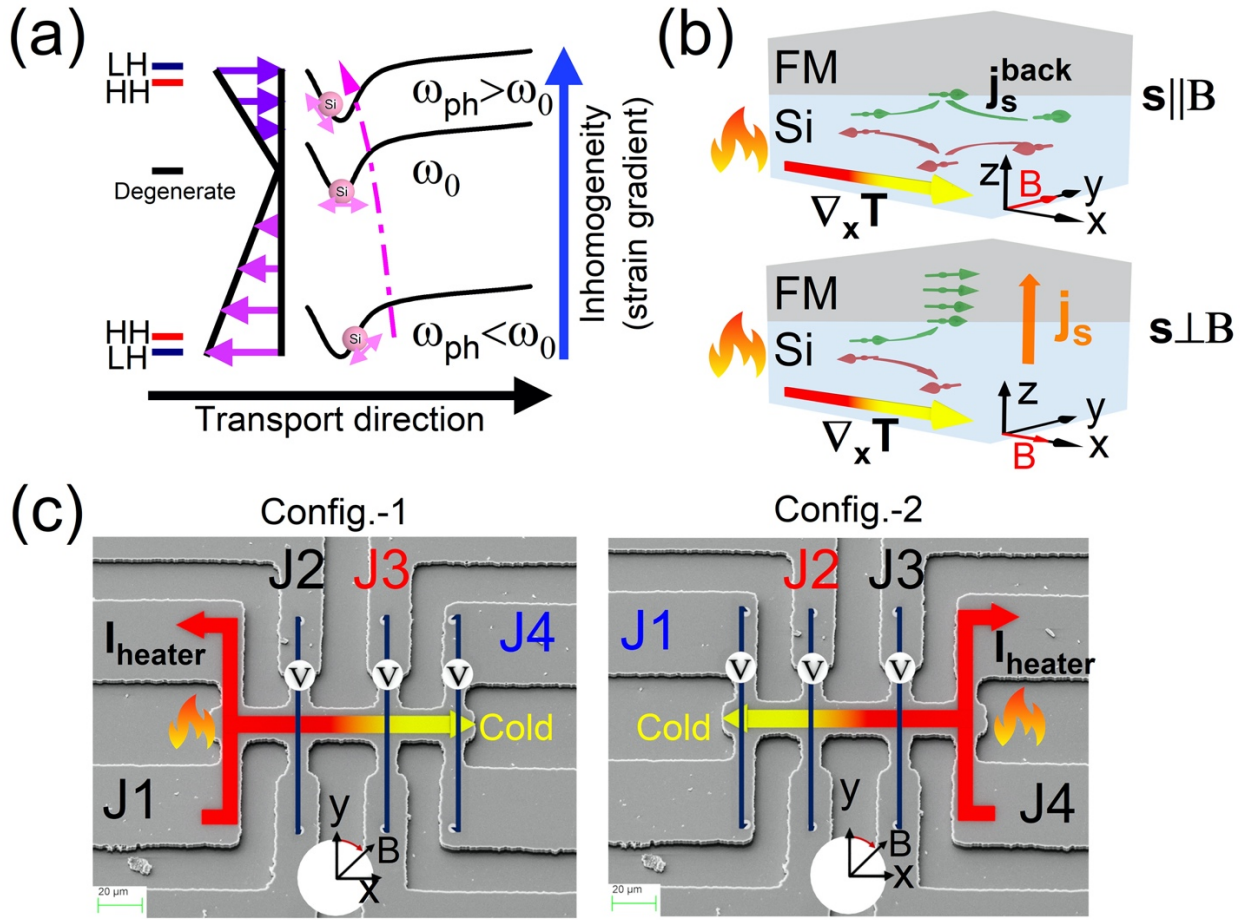

Figure 1. (a) schematic showing superposition of strain gradient and phonon transport that can give rise to wave transport in inhomogeneous medium due to change in phonon dispersion and frequency as a function of strain. The LH and HH stands for light hole and heavy hole, which give rise to flexoelectronic effect. (b) Schematic showing the angle dependent modulation of thermopower from spin-Nernst effect due to longitudinal temperature gradient ($\nabla_x T$), (c) a representative scanning electron microscope image showing the structure of the experimental device and two experimental configurations

where the ends are heated electrically by passing 2 mA current across junction J1 in Configuration 1 and J4 in Configuration 2.

Our experimental setup had four Hall junctions (J1, J2, J3, and J4) across the length of the freestanding sample composed of Pd (1 nm)/Py (25 nm)/MgO (1.8 nm)/SiO$_2$ (native)/p-Si (2 µm) (sample 1), as shown in Figure 1 (c) (Supplementary materials). The strain gradient in the sample arose due to buckling of freestanding sample from the residual and processing thermal mismatch stresses including from metal deposition[17], for which the maximum local strain could be as large as 4% based on previous studies[18]. It is noted that longitudinal inhomogeneities might also exist due to geometrical and boundary imperfections. However, a complete quantitative description of the strain gradient was not available at the time of the experiment. The transverse thermal response measurements were performed inside the Quantum Design's Physical Properties Measurement System (PPMS) at 300 K using conventional lock-in technique. The angle dependent (yx-plane) transverse $V_{2\omega}$ responses measurement was carried out in two configurations: Configuration 1- heated J1 with a 2 mA of applied current and measured response at J2 (30 µm), J3 (70 µm), J4 (100 µm), and Configuration 2- heat J4 and measured response at J3 (30 mm), J2 (70 µm), and J1 (100 µm), as shown in Figure 1 (c). The magnitude of the constant magnetic field was 1 T.

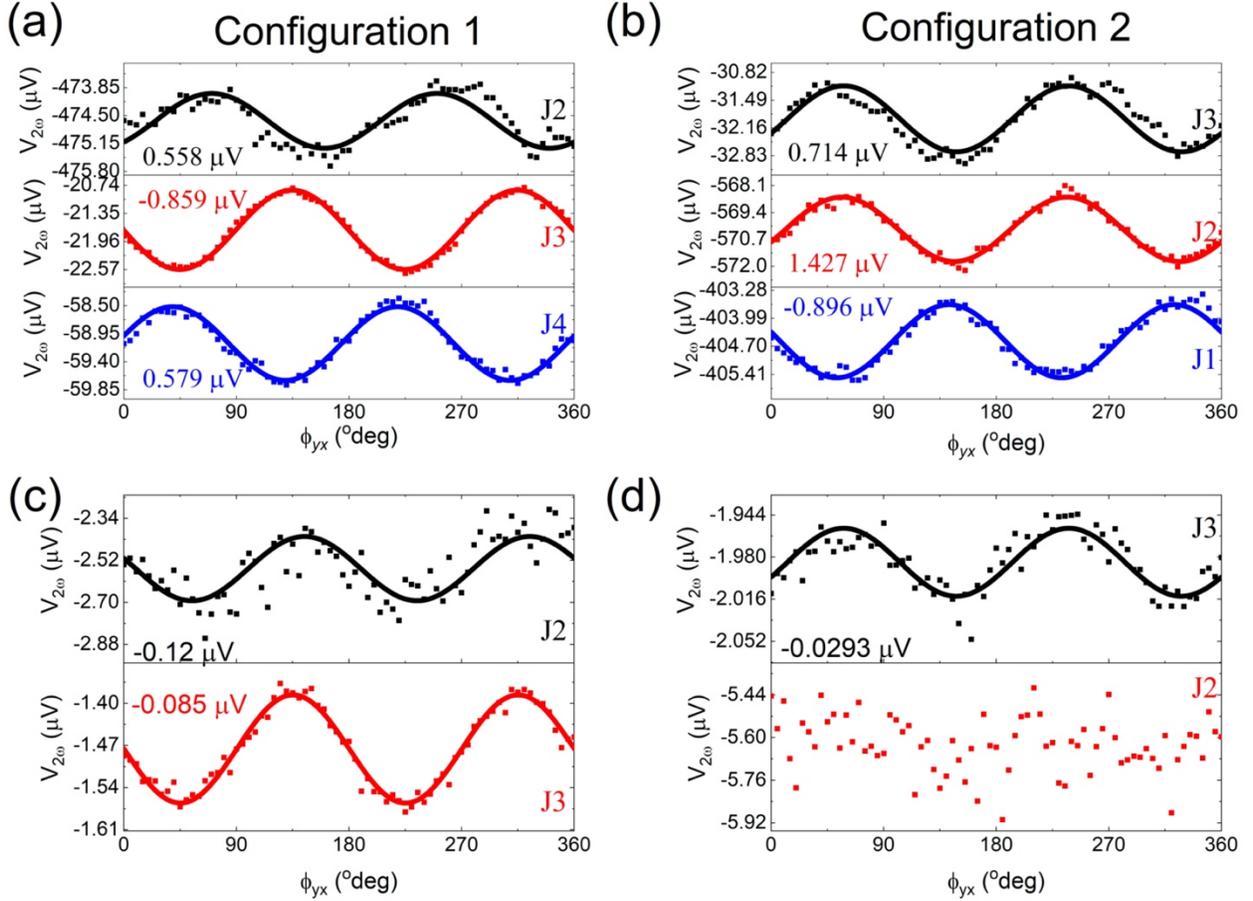

Figure 2. The angle-dependent transverse magneto-thermal transport response in Py/MgO/p-Si sample in the yx-plane at a constant applied magnetic field of 1 T (a) measured at J2, J3 and J4 in Configuration 1 and (b) measured at J3, J2 and J1 in Configuration 2. The angle-dependent transverse magneto-thermal transport response in Py/SiO$_2$ (25 nm)/p-Si control sample in the yx-plane at a constant applied magnetic field of 1 T (c) measured at J2 and J3 in Configuration 1 and (d) measured at J3 and J2 in Configuration 2.

The acquired transverse V$_{2\omega}$ responses and the amplitude of the $\sin 2\phi_{yx}$ contributions (solid line fit) for both configurations are shown in Figure 2 (a,b) and listed in Supplementary Table S1. There were only two primary contributions, having $\sin 2\phi_{yx}$ symmetry, in the angle-dependent transverse response: the planar Nernst effect (PNE)[19] response arising from the ferromagnetic layer and the expected TSNE from the non-magnetic layer (p-Si in this study)[15]. The PNE responses were measured using a control Py/SiO$_2$ (25 nm)/p-Si sample as shown in Figure 2 (c,d) and Supplementary Table S1. The 25 nm of SiO$_2$ intermediate layer extinguished the spin current as well as the interlayer coupling. The symmetry behavior of the PNE response in the control sample was consistent with that in the previously reported studies[19]. However, the PNE response in the control sample was larger than that expected for Py, which was attributed to the self-induced effect at the SiO$_2$ interface[20,21].

As compared to the PNE response in the control sample, the transverse thermal responses in the sample 1 were an order of magnitude larger. In addition, the sign of the response was not constant through the length of the sample. For example, the amplitudes of the transverse thermal responses measured at J2 and J4 were +0.5584 µV and +0.579 µV, respectively, whereas the amplitude was -0.8598 µV at J3. If these responses had originated from the PNE of the Py layer, then all three of them should have had a negative sign. Additionally, the amplitude of the PNE response in the control sample diminished as expected as we measured further away from the heat source, whereas the opposite behavior was observed in sample 1, as shown in Figure 2 (a,b) and Supplementary Table S1. For example- the responses measured at J3 and J4 were larger than J2 in configuration 1, which was contrary to the diffusive thermal transport. Additionally, the offset thermal voltages were two orders of magnitude larger in sample 1 as compared to the control sample. Based on these observations, we concluded that the measured transverse thermal responses in sample 1 were due to TSNE and Not PNE.

The estimated transverse spin-Nernst thermopower coefficient values lie between 3.5 µV/K (30 µm) to 45.4 µV/K (100 µm) as compared to 0.3 µV/K in case of W[16], as shown in Supplementary Table S1. The spin-Nernst thermopower coefficients were orders of magnitude larger than the PNE coefficient in Py (~70 nV/K)[19], which again proved that PNE was not the underlying cause of observed behavior. The measured transverse spin-Nernst thermopower coefficient magnitudes were larger than even the Seebeck coefficient of Py (-7.8 µV/K[22] to -20 µV/K[23]). The temperature profile was simulated using a COMSOL finite element model that was verified using infra-red thermal imaging, as shown in Supplementary Figures S1–S2.

The TSNE response persisted at 50 K, as shown in Supplementary Figure S3, whereas the dynamical multiferroicity, recently reported, disappeared when the temperature was reduced to 100 K due to the freezing of the optical phonons[12]. It meant that the optical phonons were not expected to give rise to the long-distance spin transport. Instead, transverse acoustic phonons[24,25] carried the spin angular momentum in this study. A careful observation of the spin-Nernst thermopower coefficients, presented in Supplementary Table S1, showed that the Hall junction closer to the heat source (30 µm) had the highest temperature gradient but the smallest coefficient, whereas the junction farthest (100 µm) from the heat source behaved in completely the opposite manner. According to Bliokh and Freilikher[14] (equation 2), topological spin-Hall effect of phonons was expected to be a function of wavelength. This suggested that the spin transport in our experiment was mediated by waves and not diffusion. Hence, the TSNE responses were, most likely, due to the topological spin-Hall effect of phonons in an inhomogeneous medium.

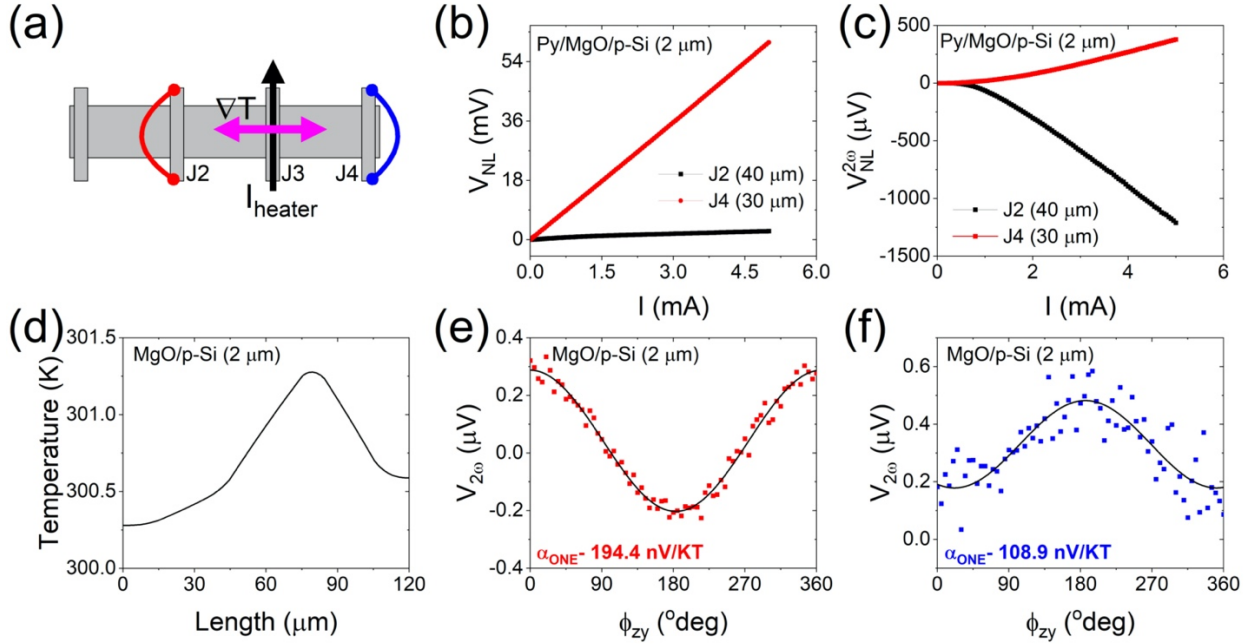

Figure 3. (a) a schematic showing the measurement scheme for non-local effects including current spreading behavior, (b) the non-local voltage ($V_{NL}$) as a function of current measured at J2 (40 μm) and J4 (30 μm) showing smaller response at J2 as compared to J4 as expected. (c) The non-local second harmonic response as a function of current measured at J2 (40 μm) and J4 (30 μm) showing larger response at J2 as compared to J4. (d) the expected temperature distribution along the length of the MgO/p-Si sample for an applied 2 mA current across the junction J3. The angle dependent second harmonic voltage measured at (e) J2 and (f) J4 showing the cosine response due to ordinary Nernst effect (ONE).

To eliminate the current spreading effect[26], we measured the non-local voltage and second harmonic response as a function of current in a Py/MgO/p-Si ( sample 2) where sinusoidal current was applied across junction J3 and responses were measured at J2 and J4 as shown in Figure 3 (a). The non-local voltage as a function of current measured at J2 (40 µm) and J4 (30 µm) showed an exponential decay of spreading current as shown in Figure 3 (b), which agreed with the van der Pauw theorem ($R_{NL} = R_{sq} e^{\frac{-\pi L}{w}}$ where $R_{sq} = \frac{\rho}{t}$, $L$- length and $w$= width of channel). The spreading current should have led to larger second harmonic response at J4 as compared to J2 since the non-local resistance was higher at J4. Whereas the second harmonic response was significantly larger at J2 as compared to response at J4 as shown in Figure 3 (c). In addition, the thermal effects due to current spreading (anisotropic thermopower) had the expected symmetry behavior of $\sin^2\phi_{yx}$ [27] and not the observed symmetry of $\sin 2\phi_{yx}$ as shown in Figure 2. This eliminated spreading current in the conducting sample as an underlying cause of thermal response reported in Figure 2.

We, then, measured the angle dependent thermal responses in a freestanding MgO/p-Si sample (sample 3) using the measurement scheme shown in Figure 3 (a). The temperature profile along the length of the sample was estimated using COMSOL and is shown in Figure 3 (d). The measured response at J2 and J4 are shown in Figure 3 (e,f), respectively. The measurement showed that the ordinary Nernst effect (ONE) was also a function of distance similar to TSNE response. The ONE coefficient was larger at 40 µm (194.4 nV.KT) away as compared to 30 µm (108.9 nV/KT) as shown in Figure 3 (e,f). This measurement on MgO/p-Si samples clearly demonstrated that topological phonons in the Si were the underlying cause of the observed behavior and the TSNE responses did not arise from Py layer in sample 1. This experiment also demonstrated that topological phonons affected both the spin and heat transport[28,29]. However, responses were an order of magnitude larger in the sample 1 with Py, which was attributed to the charge carrier transfer due to interfacial flexoelectronic effect (details in part 3).

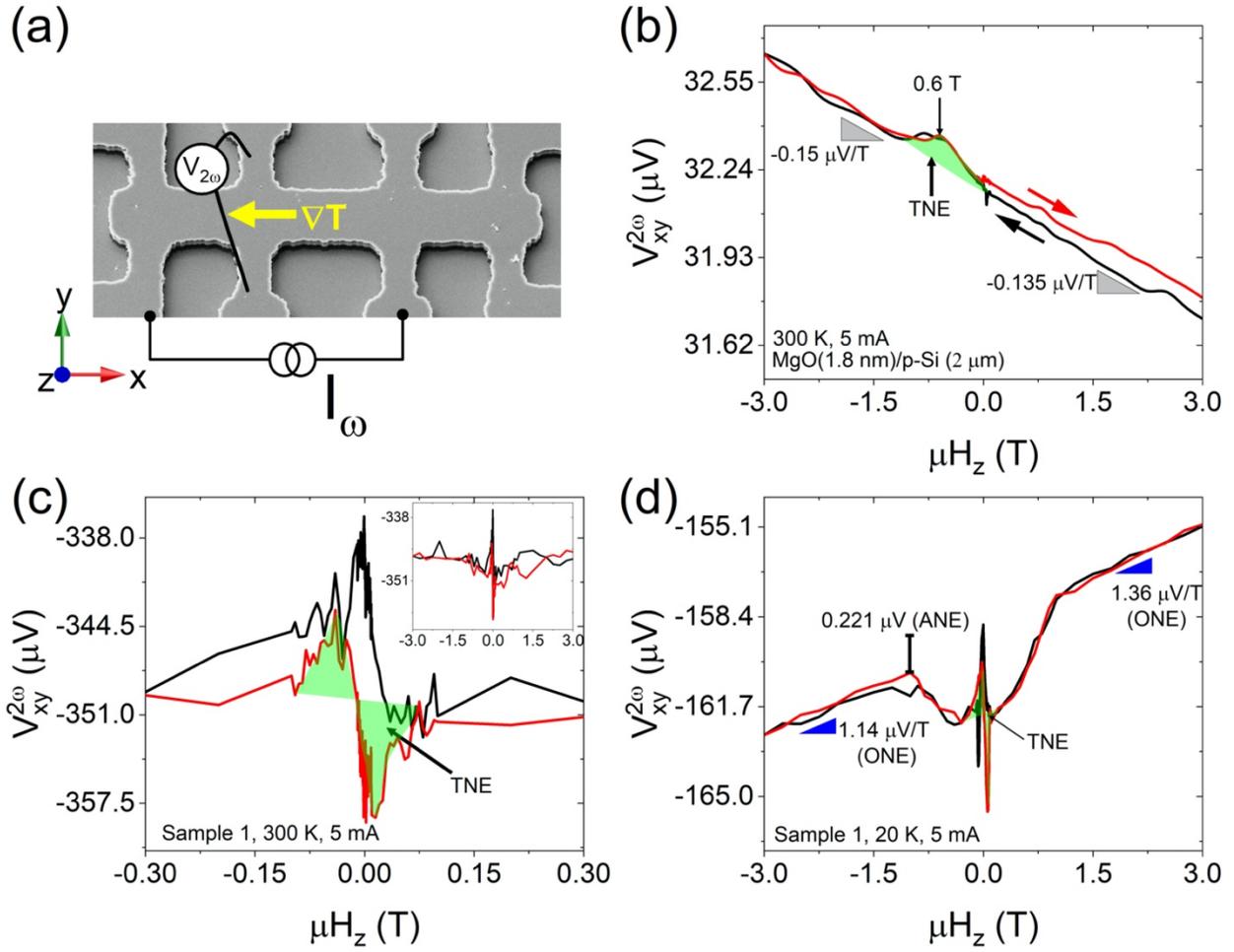

Figure 4. (a) Schematic showing the experimental setup and representative device structure to measure the second harmonic Hall voltage, which was expected to arise due to transverse thermoelectric response from self-heating. The second harmonic Hall response as a function of magnetic field from 3 T to -3 T for applied current of 5 mA and (b) in MgO/p-Si sample at 300 K, (c) Py/MgO/p-Si sample (sample 1) at 300 K and (d) at 20 K.

To uncover additional direct proof of the topological phonons in Si, we measured the transverse second harmonic Hall response as shown in Figure 4 (a). The second harmonic Hall response was measured as a function of the magnetic field from 3 T to -3

T for an applied current of 5 mA in a MgO (2 nm)/p-Si (2 µm) sample (sample 4) [30]. This sample did not have any Py layer, which allowed us to study the response from only Si but the intrinsic strain gradient in the sample was insignificant. Therefore, we applied 5 mA of current to generate large Joule heating and thermal expansion mediated strain gradient. For the sample 4, the response exhibited a negative slope and the behavior was attributed to ordinary Nernst effect (ONE) as shown in Figure 4 (b). In the Si, a negative ONE response arose due to the acoustic phonon scattering[31]. Furthermore, the magnitude of the slope was different for negative (-0.15 µV/T) and positive (-0.135 µV/T) magnetic fields, which indicated phonon skew scattering behavior; supporting previous observations[12,18]. More importantly, a bump at the negative magnetic fields (green shaded region) was observed, as shown in Figure 4 (b). This was a manifestation of the topological Berry phase in thermal transport, leading to TNE[32] from transverse phonons[31]. The TNE behavior arose due to randomized spin angular momentum of phonons. The TNE behavior was not observed in the response measured at 1 mA of the current, as shown in Supplementary Figure S4, possibly due to the smaller strain gradient. We did not observe the topological Hall effect (THE) in the Hall resistance measurement, as shown in Supplementary Figure S5. This was the first experimental evidence of a topological thermoelectric response in a non-magnetic material.

A similar measurement on the sample 1 showed a larger TNE response and an order of magnitude larger overall response as shown in the Figure 4 (c). This showed that while the response arises due to topological phonons in Si layer but it was greatly enhanced due to coupling with Py layer, most likely, due to interfacial flexoelectronic effect. The TNE response was significantly larger than ANE and ONE, which could not be discerned at 300 K as shown in Figure 4 (c). Whereas, the transverse second harmonic Hall response at 20 K was composed of distinct contributions from ONE, ANE and TNE responses as shown in Figure 4 (d). The TNE response was significantly smaller at 20 K as compared to 300 K but it did not disappear. Additionally, the positive sign of the ONE response at 20 K was attributed to the impurity scattering. Based on the TNE behavior, the acoustic phonons were expected to be the primary driver for the TNE and TSNE responses since optical phonons would freeze out at 20 K. This showed that while the response arises due to topological phonons in Si layer but it was greatly enhanced due to coupling with Py layer. This measurement also eliminated any non-linear Hall contributions[33] and established that the topological phonons were the underlying cause of the TNE response presented in Figure 4 (c,d).

We hypothesized that topological phonons arise due to inhomogeneous medium but direct estimate of the inhomogeneous strain was unavailable. The inhomogeneity due to structural inversion asymmetry and crystal inversion asymmetry in materials gives rise to non-reciprocal transport, which can be used to prove our hypothesis of inhomogeneous medium. In the transport measurements, the non-reciprocal behavior arises in inhomogeneous samples since the resistance of the sample is a function of the direction of the current, which can be described by following equation-

$$R=R_0\left[1+\beta B^2 +\gamma BI\right] \qquad (3)$$

where $R_0$, I, $\beta$, $\gamma$ and B are resistance at zero magnetic field, current, coefficient of normal magnetoresistance, coefficient of magnetoelectrical anisotropy and magnetic field, respectively. In the present study, we expected unidirectional (or nonreciprocal) magnetoresistance of phonons due to temporal magnetic moment (dynamical multiferroicity) and phonon skew scattering[34]. This behavior can also be called as dynamical magnetoelectrical anisotropy[35] due to dynamical multiferroicity of optical phonons and described by following equation:

$$\Delta R = R(I) - R(-I) \propto I^{<110>} \cdot (P_{FE}^{<112>} \times M_t^{<111>}) \quad (4)$$

The non-reciprocal transport behavior is studied using longitudinal second harmonic response since it is a quadratic function of applied ac bias. In case of sample 1, the ferromagnetic layer could lead to additional responses (such as spin-Seebeck effect, unidirectional spin-Hall magnetoresistance) that have same symmetry (cosine) in the second harmonic response. To overcome it, we fabricated a freestanding Pt (15 nm)/MgO/p-Si sample (sample 5).

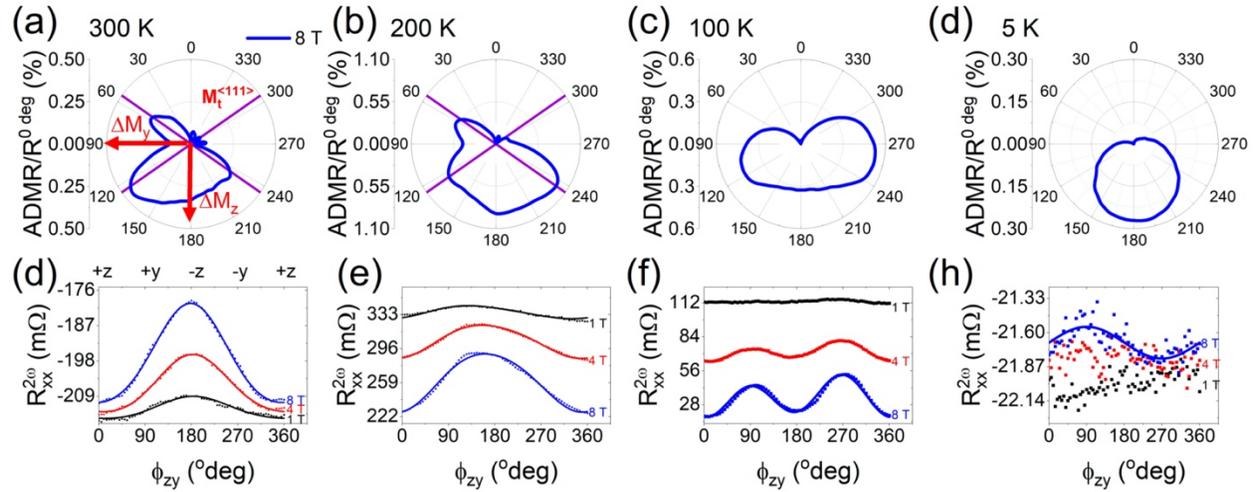

Figure 5. The angle dependent behavior in Pt (15 nm)/MgO/p-Si (2 μm) sample measured for an applied ac bias of 2 mA, magnetic field of 1 T, 4 T and 8 T rotated in the zy-plane. The angle dependent magnetoresistance behavior for the 8 T magnetic field at (a) 300 K, (b) 200 K, (c) 100 K and (d) 5 K. The angle dependent second harmonic behavior at (e) 300 K, (f) 200 K, (g) 100 K and (h) 5 K. Solid line represent curve fit.

We measured the angle dependent longitudinal resistance and second harmonic resistance for an applied ac bias of 2 mA and magnetic fields of 1 T, 4 T and 8 T as shown in Figure 5 (a-h). The sample rotation was carried out in the zy-plane ($\phi_{zy}$) for a constant magnetic field, which was also the (110) cross-sectional plane. The angle-dependent MR (ADMR) measurements exhibited behavior corresponding to dynamical multiferroicity[12] as shown in Figure 5 (a,b). The magnitude of the ADMR shows peak corresponding to the <111> crystallographic directions[12] (at 54.7° from ±z- axis) due to temporal magnetic moment along these directions, which disappeared at 100 K and below as shown in Figure 5 (c,d). It was noted that the crystallography of temporal magnetic moment cannot be attributed to the Pt since it was polycrystalline. We, then, analyzed the angle dependent longitudinal second harmonic response at 300 K, 200 K, 100 K and 5 K, which demonstrated a clear cosine behavior due to non-reciprocal transport behavior at higher temperatures as shown in Figure 5 (e-f). We fitted the responses using three contributions having sine, cosine and sine squared angular symmetries. The sine and cosine behavior were attributed to the non-reciprocal transport since temporal magnetic moment had components in both z and y directions as shown in Figure 5 (a). The sine squared response was expected to arise from longitudinal thermal transport in freestanding sample due to self-heating. Using the cosine responses, the coefficient describing the ratio of the non-reciprocal resistance was estimated to be 0.151 $A^{-1}T^{-1}$, 0.352 $A^{-1}T^{-1}$ and 0.071 $A^{-1}T^{-1}$ at 300 K, 200 K and 100 K, respectively. These values were similar to the reported value in Si FET interfaces (0.1 $A^{-1}T^{-1}$)[35-37] at 2.92 V of gate bias. While the magnitude of the non-reciprocal resistance was similar but coefficient was smaller than that reported in BiTeBr (1 $A^{-1}T^{-1}$)[36] at 2 K. Our high temperature coefficients were orders of magnitude larger than the non-reciprocal responses reported in other materials. The magnitude of the non-reciprocal response did not give us the inhomogeneous strain distribution but it did give us the quantitative information in regards to inhomogeneity especially when comparing it to the value in BiTeBr and Si FET. The measured longitudinal second harmonic response diminished as the temperature was lowered to 100 K as shown in Figure 5 (g). At 5 K, the overall response was negligible and a weak sine behavior was observed at 8 T magnetic field only as shown in Figure 5 (h). The conventional non-reciprocal response increased[36] with reduction in temperature as opposed to the observed behavior, which supported our hypothesis of phonon skew scattering mediated unidirectional magnetoresistance. Independently, the non-reciprocal response did not prove existence of the topological phonons. However, it gave us a direct proof of inversion asymmetry and inhomogeneity in the Si thin film, which was the premise of this work. Hence, when analyzed in conjunction with both the TSNE and TNE responses, it proved that the long-distance spin transport was, most likely, due to the topological phonons in an inhomogeneous medium.

    We have engineered topological phonon-mediated long-range spin transport in non-magnetic centrosymmetric material using strain gradient. Not only did we achieve long-distance spin transport, but the responses were also several orders of magnitude larger than expected, which could lay the foundation of topological and spin phononics. Previously, Cazzanelli et al.[17] demonstrated optical second harmonic response in inhomogeneously strained Si thin films. Similarly, Yang et al.[38] demonstrated large flexo-photovoltaic effect whereas Wang et al.[39] demonstrated large flexoelectronics effect in the Si thin films. Now, we have demonstrated topological phonons in

inhomogeneously strained Si thin films. These observations strongly suggest that scientific understanding of the material physics in inhomogeneously strained materials is incomplete. Our work could be a stepping stone for further research since topological behavior could potentially arise in other semiconductors and centrosymmetric insulators due to inhomogeneous strain.

**Author contributions**

The manuscript was written through contributions of all authors. All authors have given approval to the final version of the manuscript. ‡AK, RGB and PCL have equal contribution to this work.

**Acknowledgement**

The fabrication of experimental devices was completed at the Center for Nanoscale Science and Engineering at UC Riverside. Electron microscopy imaging was performed at the Central Facility for Advanced Microscopy and Microanalysis at UC Riverside. Infra-red thermal imaging microscope work was carried out at UC Santa Barbara. SK acknowledges a research gift from Dr. Sandeep Kumar.

**Supplementary information- Topological phonons in inhomogeneously strained silicon- part 1: Evidence of long-distance spin transport and unidirectional magnetoresistance of phonons**

Anand Katailiha[1‡], Ravindra G. Bhardwaj[1‡], Paul C. Lou[1‡], Ward P. Beyermann[2], and Sandeep Kumar[1,3,*]

[1]Department of Mechanical engineering, University of California, Riverside, CA 92521, USA

[2] Department of Physics and Astronomy, University of California, Riverside, CA 92521, USA

[3] Materials Science and Engineering Program, University of California, Riverside, CA 92521, USA

**Materials and Methods-**

Fabrication- All the samples are fabricated using standard nanofabrication techniques. We take Silicon on Insulator (SOI) wafer with 2 µm thick device layer. The structure of the device was patterned using photolithography and Si deep reactive ion etching (DRIE). The sample structure was made freestanding using hydrofluoric acid vapor etching. We deposited 1.8 nm of MgO layer on all the devices. Then, for sample 1 and 2, we deposited 25 nm of $Ni_{80}Fe_{20}$ (Py) thin film using e-beam evaporation. A 1 nm Pd layer was also deposited on top to protect the Py from oxidation. MgO layer was deposited to eliminate Ni or Fe diffusion as well as tunneling barrier. The width of the central beam was ~15 µm and the width of the electrodes was ~9 µm. The center to center distances of four hall junctions were- J1-J2: 30 µm, J1-J3: 70 µm and J1-J4: 100 µm. The samples 3 and 4 were without any metal layer and were expected to have smaller strain gradient and inhomogeneity. The sample 5 had 15 nm of Pt layer on top of MgO layer. Even though Pt layer was thinner but residual stresses due to Pt would still be large due to higher melting point than Py.

The resistivity of the device layer Si was 0.001–0.005 $\Omega$cm. The magnitude of the applied current across both the Py and p-Si layers was expected to be ~50% in sample 1. In sample 5, the resistivity of the Pt and p-Si layers were $2.52\times10^{-7}$ $\Omega$-m and $1.1\times10^{-5}$ $\Omega$-m, respectively, and ~75% of the current will pass through the p-Si layer.

**Supplementary Figures-**

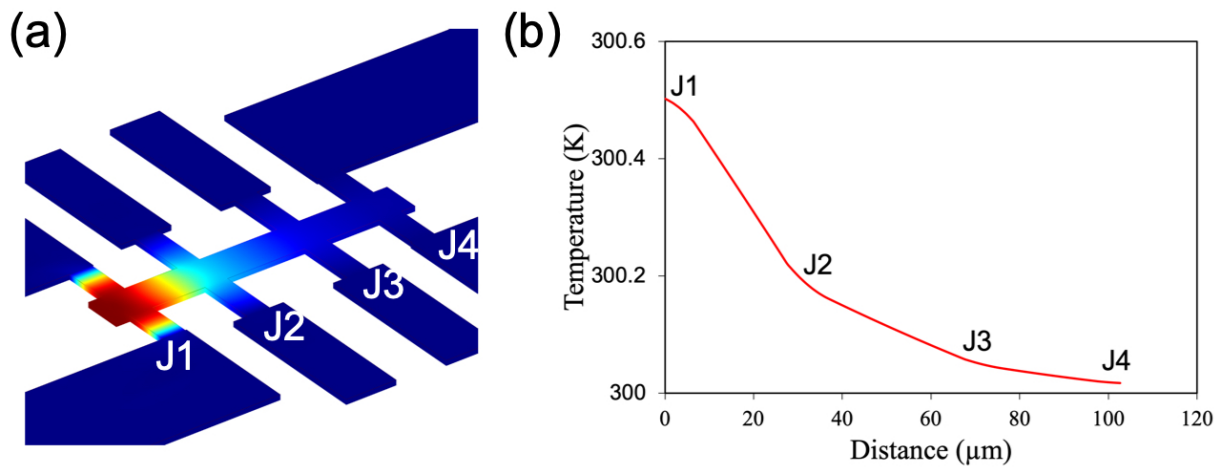

Supplementary Figure S1. (a) The expected temperature distribution across the length of the specimen was estimated using COMSOL simulation at 2 mA of the current and (b) the longitudinal temperature profile for an applied heating current of 2 mA. The temperature gradients were estimated to be 10.59 K/mm, 3.46 K/mm and 1.05 K/mm between junctions J1-J2, J2-J3 and J3-J4, respectively. The spin-Nernst thermopower coefficient values were calculated using $S_{xy} = \frac{\Delta V_{xy}}{w \cdot \nabla_x T}$, where $\Delta V_{xy}$, $w$ and $\nabla_x T$ were amplitude of transverse thermal voltage, width and temperature gradient, respectively.

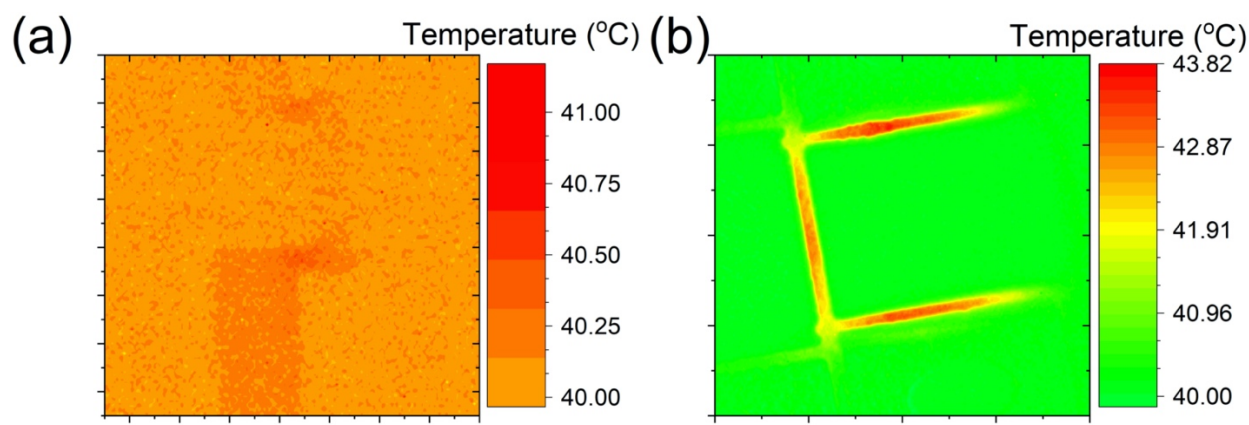

Supplementary Figure S2. The infra-red thermal imaging microscope image showing the temperature profile (a) for 4 mA of the longitudinal current in the Py/MgO/p-Si sample. It did not show any temperature increase for 4 mA of the current, which was attributed to heat loss to the air. (b) Then, we used a Py/MgO/p-Si composite sample device with back side etched similar to the devices used in Reference's 17 and 33. The heat loss to the air in this sample will be insignificant because there is no heat sink underneath the sample region. We estimated a thermal conductivity of 85-90 W/mK using a finite element model in COMSOL, which is consistent with the values reported over the years and the same value was used to estimate the temperature profile in our sample. The temperature of the stage was 40°C in the measurement.

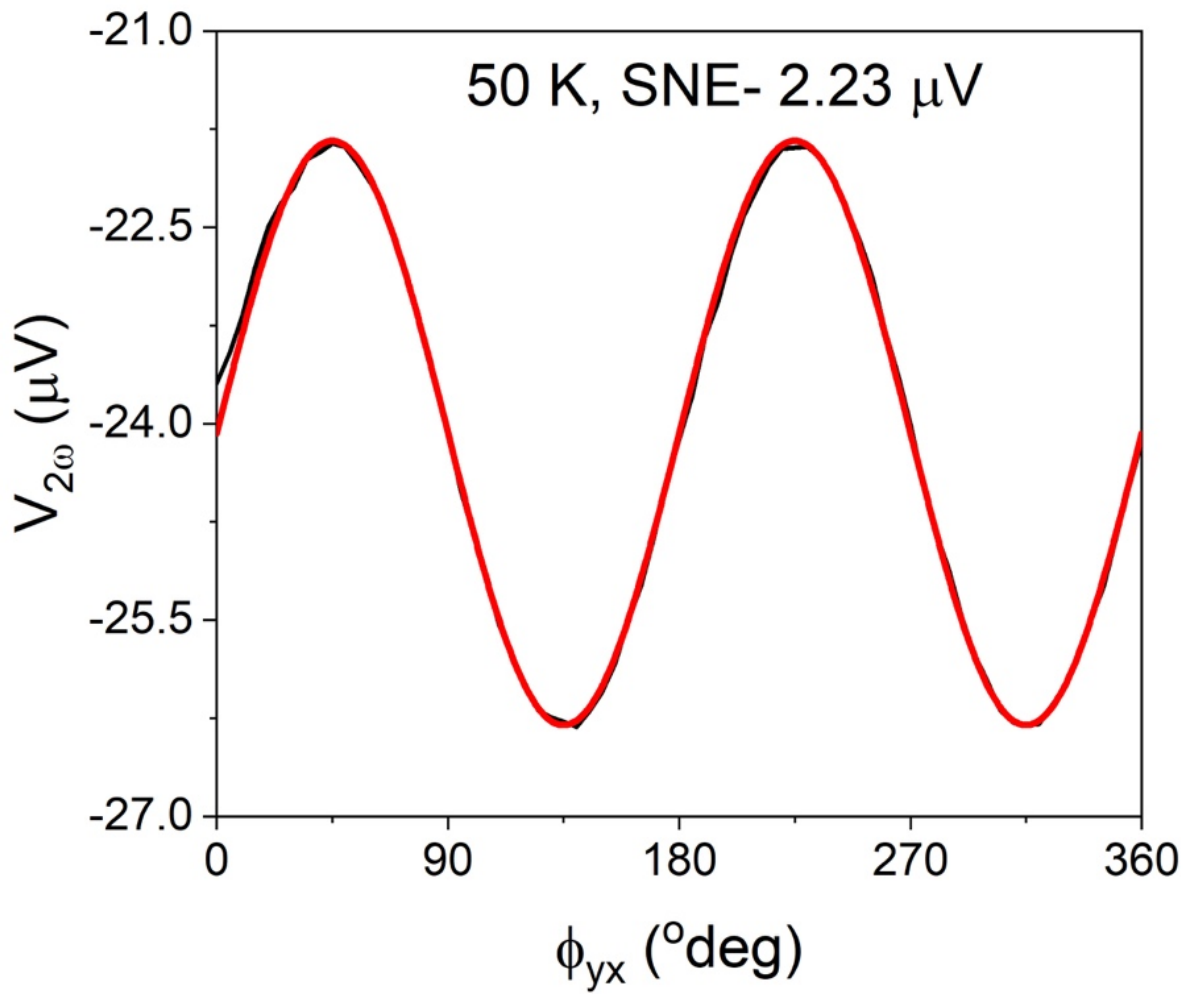

Supplementary Figure S3. The angle-dependent transverse thermoelectric response measured at junction J2 for an applied magnetic field of 4 T at 50 K. The heating current was applied across junction J1.

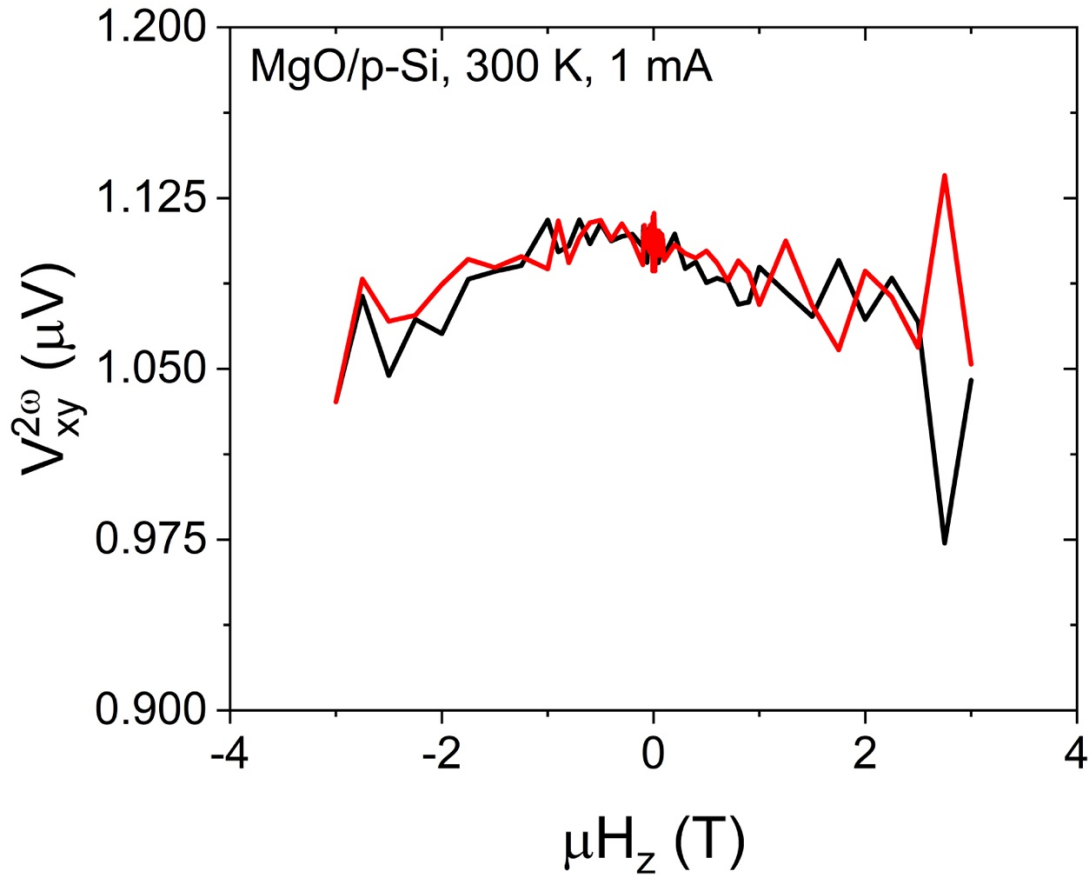

Supplementary Figure S4. The second harmonic Hall response in MgO (1.8 nm)/p-Si (2 µm) sample (sample 4) as a function of magnetic field from 3 T to -3 T at applied current of 1 mA. The residual stresses in this sample were small as compared to the sample 1 having Py layer since the metal layer deposition (sample 1) enhanced the strain gradient due to thermal mismatch stresses. As a consequence, the TNE behavior was not observed in this measurement on sample 4.

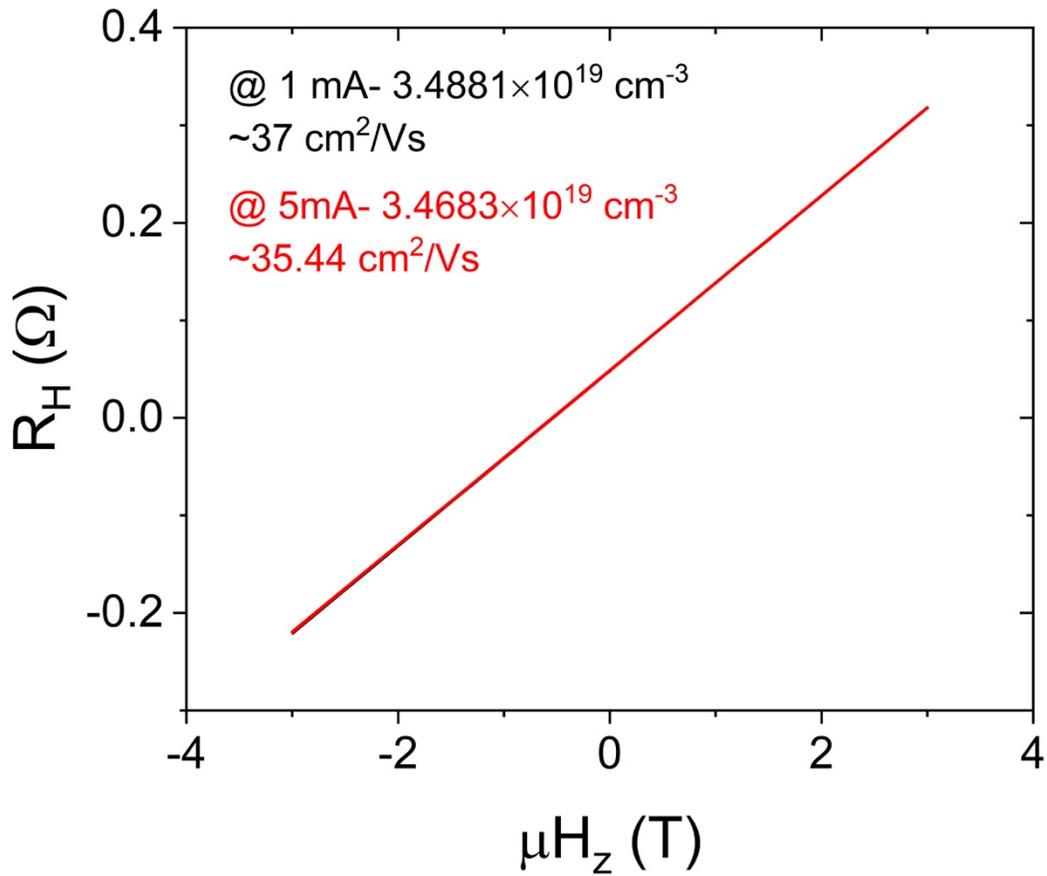

Supplementary Figure S5. The Hall resistance measurement in MgO (1.8 nm)/p-Si (2 μm) sample (sample 4) for an applied current of 1 mA and 5 mA showing hole mediated transport behavior. There was no topological Hall effect (THE) response corresponding to the TNE response in Figure 4 (b).

**Supplementary Table-**

Supplementary Table S1. The magnitude of the angular modulation in the transverse $V_{2\omega}$ response for the direction-dependent thermal transport and the resulting symmetry behavior of the $V_{2\omega}$ response.

| | Junction | J1 | J2 | J3 | J4 |
|---|---|---|---|---|---|
| **Ni₈₀Fe₂₀/MgO/p-Si** | TSNE/PNE response | Heater (-$\nabla_x T$) | 0.5584 µV | -0.859 µV | 0.579 µV |
| | Magnitude of Coefficient | | 3.51 µV/K | 16.54 µV/K | 36.79 µV/K |
| | TSNE/PNE response | -0.896 µV | 1.427 µV | 0.714 µV | Heater (+$\nabla_x T$) |
| | Magnitude of Coefficient | 45.4 µV/K | 27.5 µV/K | 5.63 µV/K | |
| **Control Ni₈₀Fe₂₀/SiO₂ (25 nm)/p-Si** | PNE response | Heater (-$\nabla_x T$) | -0.12 µV | -0.085 µV | Not measured |
| | PNE response | Not measured | negligible | 0.0293 µV | Heater (+$\nabla_x T$) |